# White LED-based photocatalytic treatment using recoverable cobalt ferrite nanoparticles

Naresh Prajapati[a], Manoj Kumar[b], Vidit Pandey[c,d], Sandeep Munjal[e], Himanshu Pandey[a,*]

[a] Condensed Matter & Low-Dimensional Systems Laboratory, Department of Physics,
Sardar Vallabhbhai National Institute of Technology, Surat 395007, India
[b] Department of Physics and Materials Science and Engineering,
Jaypee Institute of Information Technology, Noida 201309, India
[c] Department of Physics, Aligarh Muslim University, Aligarh-202002, India
[d] Research & Innovation Unit, SAIR, Aligarh-202001, India
[e] National Forensic Sciences University, Goa-403401, India

**Abstract:** Contamination of freshwater sources has been alarming due to the widespread use of toxic chemicals in various industries. Advanced oxidation processes (AOPs) such as photocatalysis are widely explored to tackle such problems. In photocatalysis, highly oxidative species such as hydroxyl radicals (*OH) are produced with the help of some semiconductor photocatalysts and light. A photocatalyst decomposes these toxic organic compounds in the presence of light. Spinel ferrite ($M$Fe$_2$O$_4$, $M$ = Co, Ni, Cu, Zn, *etc*.) materials are an important candidate as a photocatalyst due to their semiconducting behaviour and narrow optical bandgap. In this work, we have synthesized cobalt ferrite (CoFe$_2$O$_4$) nanoparticles using the sol-gel method and subsequently annealed at 500°C. The nanoparticles are characterized using X-ray diffraction, scanning electron microscopy, Raman, and Infrared spectroscopy for structural analysis. The band gap of the material is evaluated using UV-visible spectroscopy. The photocatalytic activity of the material is investigated using methyl orange and methylene blue aqueous solutions as a model dye and a low-power white LED as a light source. The material could decompose 95 % of the dye after 150 minutes of irradiation. Adding hydrogen peroxide further improves the decomposition rate, with over 90 % decomposition achieved within 90 minutes.

*Corresponding author
hp@phy.svnit.ac.in (H. Pandey)

# 1. Introduction

In the present scenario, the importance of the catalyst is unimaginable. Catalysts are widely used in various industries, such as pharmaceuticals, oil refineries, petrochemicals, textiles, and food industries [1]. The most important property of any catalyst is to accelerate the reaction rate of any chemical process; catalysts are not consumed and can be reused repeatedly for the same process. Nowadays, it is becoming a trend to use catalysts to cure many environmental problems like air and water pollution [2,3]. Many research articles are being published in which varieties of nanocatalysts are studied to tackle these environmental problems [4,5]. Among nanocatalysts, including metal oxides, metal nitrides, metal carbides, metal sulphides, *etc*., metal oxides are the most stable, abundant, cheap, and easy to produce[6–8]. Their chemical and physical properties can be manipulated by controlling morphology, size, annealing temperature, and doping [9–11]. These materials are widely studied for their versatile applications in the field of photocatalysis.

In many countries, the contamination of freshwater resources has become a critical environmental issue. This problem is primarily driven by the extensive use of toxic chemicals across various industries, including textiles, agriculture, and manufacturing [12–14]. These chemicals often find their way into natural water bodies, leading to severe environmental and health consequences. As a result, there is a pressing need for effective treatment methods to address the growing water pollution problem. Among the various water treatment technologies available, advanced oxidation processes (AOPs) can be considered as most promising and efficient solutions for combating chemical contamination in water. AOPs are a set of chemical treatment methods that generate highly reactive species, such as hydroxyl radicals (*OH), which are capable of breaking down and neutralizing a wide range of organic and inorganic pollutants in water. These hydroxyl radicals are potent oxidants that can rapidly degrade harmful substances, converting them into non-toxic by-products such as water and $CO_2$. To enhance the generation of hydroxyl radicals and accelerate the oxidation reactions, AOPs often involve the use of catalysts. These catalysts can be ultraviolet (UV) light, ozone ($O_3$), or even hydrogen peroxide ($H_2O_2$). Using these catalysts helps improve the process's efficiency, making it possible to treat large volumes of contaminated water effectively. AOPs have been found to be particularly effective in treating recalcitrant pollutants that are resistant to conventional treatment methods [15,16]. This method uses semiconductors and light sources to decompose organic pollutants. The AOPs are environmentally friendly and can oxidize these toxic pollutants into non-toxic waste that can be discarded in the environment. Photocatalysis is one such example of AOPs. Photocatalysis requires a highly stable and non-reacting semiconductor photocatalyst with an appropriate bandgap. Ferrite nanomaterials have been widely studied for their photocatalytic activity attributed to their unique structure and stability under various reaction environments [17–19].

One of the significant pollutants discharged from industries is dye-contaminated wastewater. Reports show that thousands of types of dyes have been manufactured and sold, with a net global consumption of around one million tons [20]. These compounds are cationic and anionic, based on molecular structure and how they dissociate once dissolved in water. These molecules are highly stable due to the presence of tertiary amine and carboxylic groups in the structure [21]. Upon direct or indirect exposure and ingestion, a profound impact on living animals and the ambient environment has been reported [22]. Being so harmful and dangerous to the biosphere, effectively removing these dyes from wastewater remains a major challenge.

Ferrites are a type of metal oxide that is magnetic in nature. Ferrites are widely investigated as catalysts in many reactions, including photocatalysis [17–19,23,24]. Most ferrites are semiconductors having a bandgap in the visible region [18,25]. The properties of these ferrites can be manipulated by doping with different cations or forming composite phases with other metal oxides [26]. Based on their crystal structure, ferrites have been classified into perovskite, spinel, and hexaferrites. Spinel ferrites are expressed using the formula $M$Fe$_2$O$_4$ ($M$ = Ni, Co, Cu, Zn, Mn). Introducing these metals in spinel ferrites improves their other properties while retaining strong magnetism. So, these materials are the first choice for the photocatalysis of wastewater, in which catalysts can be easily separated magnetically after the reaction [27,28]. Thus, spinel ferrites are emerging as important catalysts among researchers due to their promising performance in the photocatalytic treatment of dye-contaminated wastewater.

Among these spinel ferrites, cobalt ferrite (CoFe$_2$O$_4$ or CFO) magnetic nanoparticles are the centre of attention because of their magnetic nature, high structural stability, and narrow optical bandgap [29,30]. It possesses a cubic inverse spinel structure in which $Fe^{3+}$ ions occupy all the tetrahedral sites (A-site), whereas octahedral sites (B-sites) are shared among $Co^{2+}$ and $Fe^{3+}$ ions. These properties can be manipulated by opting for different preparation methods. CFO nanoparticles can be prepared by a variety of synthesis methods, such as sol-gel [31], hydrothermal method [32], co-precipitation [25], solid-state method [33], and solution combustion method [34]. In these methods, the annealing temperature was the most critical factor in determining various properties of the CFO nanoparticles. In recent times, photocatalytic removal of dyes from wastewater using ferrite nanoparticles has drawn the attention of many researchers, opening a new field in wastewater treatment [35].

The critical characteristic of any photocatalyst is its bandgap, which must be in the visible region. Most researchers have used $TiO_2$ and ZnO as photocatalysts, known as first-generation photocatalysts [36,37]. These materials have wider bandgap and work under UV lights only. This is the main hurdle for commercializing this technique due to the high production cost per watt of UV light. The solar spectrum also contains only 4% of its energy in the UV range, and the usage of solar light has many limitations. Therefore, as mentioned earlier, most spinel ferrites are semiconductors having bandgap in the range of (~ 1.2-2.8 eV) [17,38]. Due to this property, spinel ferrites can degrade different organic pollutants using visible light. Many researchers have explored ferrite nanoparticles for the photocatalytic removal of contaminants. People have used $MFe_2O_4$ (M = Cu, Co, Ni, Zn) to degrade various dyes such as Rhodamine B, Congo red, crystal violet, methylene blue, and methyl orange at different pH [39]. Though these studies have shown remarkable results, some drawbacks include low pH requirement, slow degradation kinematics, external cooling, and a UV lamp or high-power Xe-lamp requirement. Again, these restrictions impede the possibility of commercialization of this technique.

In this work, the spinel CFO nanoparticles were synthesized using the tartaric acid-based sol-gel method. These nanoparticles were annealed at different temperatures to optimize their structural and optical properties and then characterized for their structural, morphological, vibrational, and optical properties using various experimental techniques. The photocatalytic activity of synthesized nanoparticles was investigated through the degradation of two different anionic (methyl orange, *i.e.*, MO) and cationic (methylene blue, *i.e.*, MB) dyes. A low-power 12-Watt (W) white light emitting diode (LED)-based light was employed as an irradiation source. The degradation parameters were calculated based on the estimation using first-order reaction kinetics. Based on analyses and outcomes, a schematic model has been proposed to understand the photocatalytic mechanism.

## 2. Experimental Techniques

### 2.1. Synthesis of CFO nanoparticles

CFO nanoparticles were synthesized using the tartaric acid-based sol-gel method as illustrated in Fig. S1 [40–42]. For the synthesis of CFO nanoparticles, we used Cobalt nitrate hexahydrate [(Co(NO$_3$)$_3$·6H$_2$O] from Merck (99%)] and ferric nitrate nonahydrate [(Fe(NO$_3$)$_3$·9H$_2$O), from Merck (99%)] as a precursor. A stoichiometric amount of nitrates was dissolved in distilled water in different glass beakers and stirred till it became clear. The stoichiometric amount of chelating agent tartaric acid [(C$_4$H$_6$O$_6$), from Merck (99%)] was (3:4 molar proportion with nitrates) dissolved in distilled water in another separate beaker. All these solutions were mixed in a larger glass beaker and placed on a magnetic stirrer with a hot plate. The final solution was stirred for 30 min for the homogenization. After that, 1 M solution of Ammonium hydroxide was added dropwise to neutralize (pH ~ 7) the final solution. The solution was continuously heated and stirred at 60-70 °C for 5 hours. At last, most water evaporated, and the solution turned into a dark, viscous gel. On further heating, the get was burnt to obtain porous black xerogel. This xerogel was ground in an agate mortar pestle for a 3 hours and converted into fine black powder. This fine powder was divided into four equal proportions and annealed in the tube furnace at four different annealing temperatures (400-550 °C) for 2 hours at the heating rate of 5 °C min$^{-1}$. Finally, prepared samples were labelled and stored in glass vials. Using different characterisation techniques, these samples were further characterized for studying structural, optical, vibrational, thermal, and photocatalytic activity. The product formed can be analysed with the following chemical reaction:

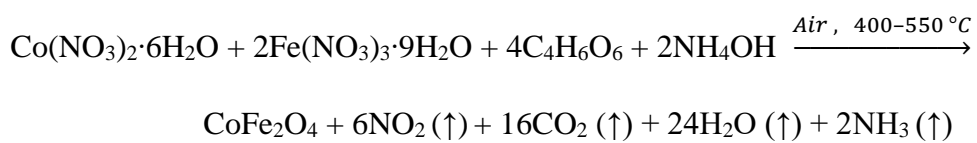

$$Co(NO_3)_2·6H_2O + 2Fe(NO_3)_3·9H_2O + 4C_4H_6O_6 + 2NH_4OH \xrightarrow{Air,\ 400-550\ °C}$$

$$CoFe_2O_4 + 6NO_2 (\uparrow) + 16CO_2 (\uparrow) + 24H_2O (\uparrow) + 2NH_3 (\uparrow)$$

### 2.2. Characterization

The structural property was investigated by X-ray diffraction (XRD) using the Rigaku Miniflex-II diffractometer with Cu $K\alpha$ radiation in the $2\theta$ range of 15˚-75˚ with the step size of 0.02˚. Rietveld analyses of the XRD patterns were carried out using the FullProf suite software. The morphological characterization of the nanoparticles was performed by scanning electron microscope (SEM) (Hitachi S-3000N). Fourier transform infrared (FTIR) analysis was carried out for vibration properties and identification of chemical bonds using JASCO FT/IR-6600 type A in the incident wavelength range of 400-4000 cm$^{-1}$. Raman spectrographs were carried out using the Emoptic Pro-Raman instrument. Optical properties and bandgap were studied using UV-visible (UV Vis.) DRS analysis by Thermo Fisher Evolution 260 set-up.

**2.3. Photocatalytic activity measurement**

The photocatalytic activity of synthesized CFO nanoparticles was studied using the degradation of aqueous solutions of MO and MB under visible light. The experiment was performed at room temperature and neutral pH. Based on structural, morphological, and optical investigations, the 500 °C annealed CFO nanoparticles were employed in photodegradation activity. A 10 mg/L MO and MB stock solution was prepared in the distilled water. In this experiment, two different cases were taken. First, we studied the photocatalytic activity of CFO nanoparticles without using any oxidizer. In the second case, we added hydrogen peroxide ($H_2O_2$) as an oxidizer to improve the reaction speed. In the first case, 50 ml of dye solution was taken in a borosilicate glass beaker, and 25 mg (500 mg/L) of CFO nanoparticles were dispersed in it. It was stirred in a dark for 30 minutes to attain the adsorption-desorption equilibrium state. Consequently, the suspension was irradiated by the visible light source (12 W White LED light) under constant stirring for 150 minutes. Aliquots of 3 ml were collected from the solution at specific time intervals, and the catalyst was removed magnetically from the liquid mixture. Similarly, in the second case, we added calculated amount of 0.05 ml $H_2O_2$ (30 % w/v, ThermoFisher Scientific, India) solution to the mixture before stirring in the dark. . The final mixture was stirred in a dark for 30 minutes to attain the adsorption-desorption equilibrium state. Subsequently, the light source was turned on, and the remaining procedure was followed, as done in the first case. The UV-Vis absorption spectrophotometer was used to measure the dye concentration in the collected samples. The degradation percentage was calculated from the following equation,

$$\eta = \left(\frac{C_0 - C_t}{C_0}\right) \times 100 \qquad (1)$$

where $C_0$ and $C_t$ are the concentrations of dye solution at the beginning and after the time interval (*t*).

The scavenger test was performed using Isopropyl alcohol–IPA (99 % Maxxic Lab, India). For the determination of active species, 0.05 ml of IPA was added to the mixture. This test was performed for both cases and on both dyes. For the recyclability test, the photocatalyst was recovered magnetically and washed several times with distilled water and IPA. After washing, the nanoparticles were dried at 100 °C in hot air oven for 3 hours. This process was repeated for five consecutive cycles, and degradation efficiency was calculated using Eq. (1)

## 3. Results and Discussions

### 3.1. XRD analyses

The room temperature XRD patterns of synthesized CFO nanoparticles were recorded for all four samples annealed at 400, 450, 500, and 550 °C and were labelled as $T_1$, $T_2$, $T_3$, and $T_4$. The XRD patterns of CFO nanoparticles [Fig. 1(a)] shows all the significant peaks belonging to the cubic phase with $Fd\bar{3}m$ space group symmetry. All peaks perfectly matched the standard data from JCPDS card no. 22-1086. Major diffraction peaks of CFO were observed around 18.34°, 30.17°, 35.53°, 37.17°, 43.19°, 53.6°, 57.1°, and 62.72° corresponding to (111), (220), (311), (222), (400), (422), (511), and (440) Miller planes. For the samples annealed at 550 °C, peaks from the $Fe_2O_3$ phase were detected, as shown in Fig. 1(a). The peak position in XRD patterns is related to the crystal structure of the material and all the space group symmetry. For further confirmation, a detailed Rietveld analysis of XRD patterns was performed using the FullProf suite [43].

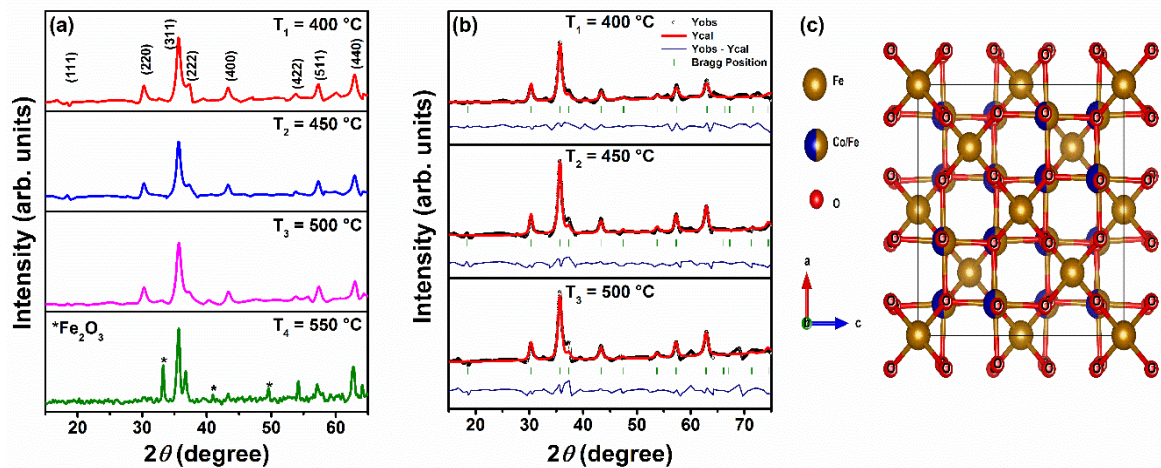

**Fig. 1.** (a) XRD patterns of synthesized CFO nanoparticles annealed at different temperatures (b) Results of Rietveld refinement of XRD pattern. (c) Cubic inverse spinel structure of CFO nanoparticles.

The refinement of XRD patterns was initiated using the $Fd\bar{3}m$ space group, and values of cell parameters were taken from the previous studies. The *Thomson-Cox-Hasting Pseudo voigt* function was selected for profile fitting, and the background was fitted using a sixth-order polynomial. All fitting parameters were refined individually to obtain a decent fit until the value of $\chi^2$ (Goodness of fit) was minimized. The final refinement results of all XRD patterns, along with the Bragg position, are shown in Fig. 1(b). Calculated intensities and experimentally observed data are represented with solid red lines and black circles, respectively. The small green vertical line below each peak represents the Bragg position belonging to the $Fd\bar{3}m$ phase of CFO, and a solid blue line at the bottom of the plot represents the difference between calculated and observed data. These plots confirmed that synthesized CFO nanoparticles possess cubic inverse spinel structure with $Fd\bar{3}m$ space group, as shown in Fig. 1(c). The refined lattice parameters are tabulated in Table 1.

**Table 1**
Rietveld refined lattice parameters (*a*, *b*, and *c*) and corresponding angles (*α*, *β*, and *γ*) of inverse cubic spinel cobalt ferrite.

| Annealing Temperature (°C) | Lattice parameters | | | |
|---|---|---|---|---|
| | $a=b=c$ (Å) | $α=β=γ$ | V (Å³) | Density $ρ$ (g/cm³) |
| 400 | 8.357 ± 0.002 | 90° | 583.69 ± 0.42 | 5.340 |
| 450 | 8.360 ± 0.001 | 90° | 584.26 ± 0.21 | 5.335 |
| 500 | 8.355 ± 0.002 | 90° | 583.23 ± 0.42 | 5.344 |
| 550 | 8.377 ± 0.003 | 90° | 587.68 ± 0.63 | 5.304 |

The crystallite size was calculated using two different methods: Scherrer's method and the Williamson-Hall (WH) plot method. In Scherrer's method, the broadening is considered due to the crystallite size only, whereas in the WH plot method, crystallite size and strain are considered for the broadening of the XRD peak. Considering all these factors, the crystallite size was estimated using the full width at half maxima (FWHM) values obtained by fitting XRD peaks using the Gaussian function. This process was repeated for all four samples.

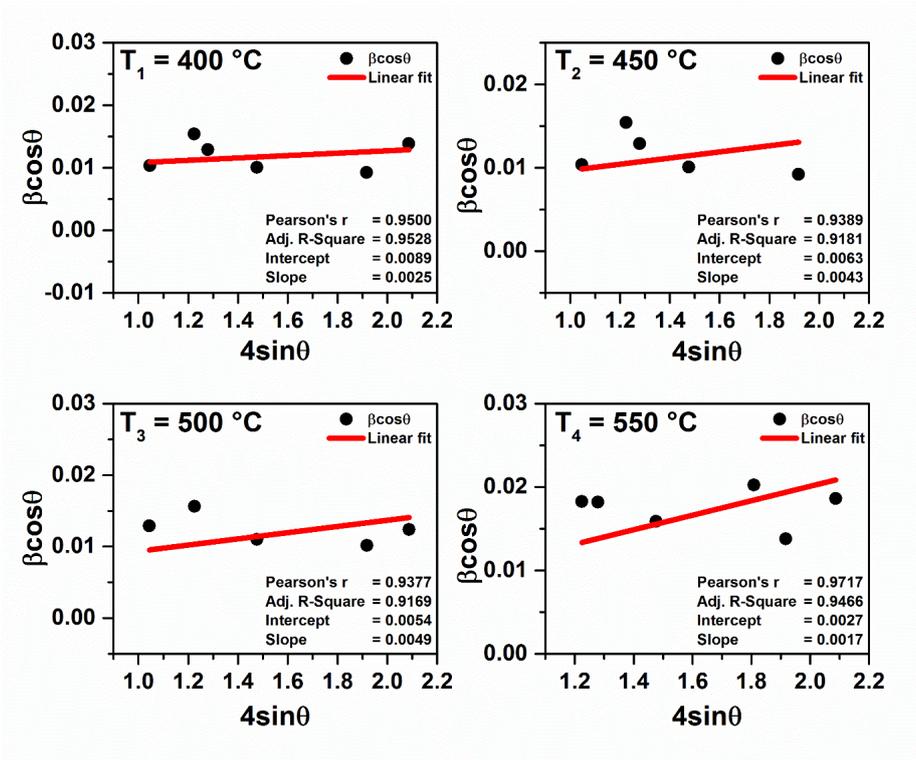

**Fig. 2.** WH plots for the calculation of crystallite size and strain.

The formula to calculate crystallite size (*D*) is known as Scherrer's formula and is given by

$$D = \frac{k\lambda}{\beta \cos\theta} \quad (2)$$

where *k* is the shape factor, *λ* is the X-ray wavelength (here, *λ* = 0.15425 nm), *β* is *FWHM* measured for a particular peak, and *θ* is the Bragg angle. The crystallite size was also calculated using a WH plot (Fig. 2) using the FWHM data obtained from the Gaussian fitting of all major peaks. In the WH plot method, the broadening of the peak is assumed due to the crystallite size as well as the strain developed in nanoparticles. Thus, the Eq. (2) can be re-written as by including the broadening effect due to the strain,

$$\beta = \beta_{size} + \beta_{strain} \quad (3)$$

Substituting values for both terms,

$$\beta = \frac{k\lambda}{D\cos\theta} + 4\eta\tan\theta \quad (4)$$

The broadening due to crystallite size and strain can be calculated from the first and second terms in Eq. (4). By taking *x* = *4sinθ* and *y* = *βcosθ*, we can plot a straight line between these two, with a slope value equal to *η* and intercept at *kλ/D*. The slope value represents strain, and we can easily calculate crystallite size *D* from the intercept value. The crystallite size value obtained from both methods is compared below in the graph, as shown in Fig. 3(b). It was found that the crystallite increases with the increase in the annealing temperature. This may be caused by the smaller crystallites fusing together to grow more prominent at higher annealing temperatures. On comparison between Scherrer's and WH plot methods, it was found that Scherer's method yields law crystallite size compared to the WH plot method. This may be because we have also included the effect of strain in the second one.

In the strain analysis, it was observed that the strain was decreasing with increments in the annealing temperature. This can be explained by the crystal's growth due to rising annealing temperature, which can form dense and fine compact crystals, reducing the strain in the structure. The same explanation also justifies the increasing crystallite size, which may eventually decrease the strain in nanocrystals.

Another parameter, the dislocation density ($\delta$), was also calculated from the size-strain analysis. It is defined as the length of dislocation lines per unit volume and expressed as,

$$\delta = \frac{1}{D^2} \tag{5}$$

The value of dislocation density was calculated using crystallite size obtained from both methods discussed earlier. The dislocation density decreased with an increase in the annealing temperature. The crystallite size, strain, and dislocation density are summarized in Table 2.

**Table 2**
Crystallite size (*D*), dislocation density ($\delta$), and strain ($\eta$) were calculated using *FWHM* parameters.

| Annealing Temperature | Scherer's method | | WH Plot Method | | |
|---|---|---|---|---|---|
| | *D* (nm) | $\delta \times 10^{-3}$ (nm$^{-2}$) | *D* (nm) | $\delta \times 10^{-3}$ (nm$^{-2}$) | $\eta \times 10^{-3}$ (%) |
| $T_1$ = 400 °C | 11.99 | 6.96 | 15.6 | 4.11 | 2.53 |
| $T_2$ = 450 °C | 12.03 | 6.91 | 22.04 | 2.06 | 2.37 |
| $T_3$ = 500 °C | 12.53 | 6.37 | 25.71 | 1.51 | 1.97 |
| $T_4$ = 550 °C | 21.39 | 2.19 | 51.42 | 0.38 | 1.74 |

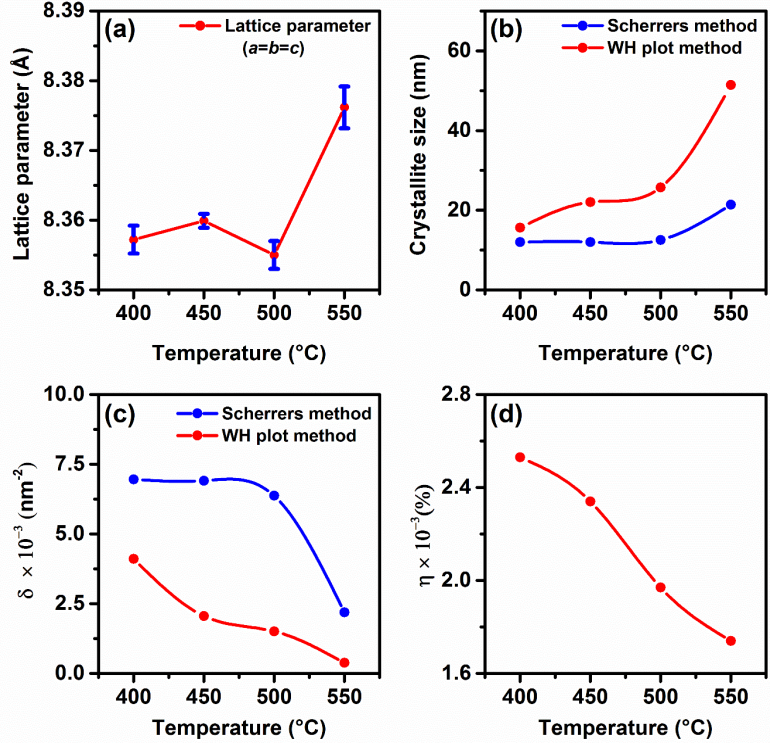

**Fig. 3.** (a) Variation in lattice parameters with change in the annealing temperature. Comparison of (b) crystallite size (*D*), (c) Dislocation density (*δ*), and (d) strain (*η*) obtained using two different methods.

The electron density distribution of the most intense plane (311) is shown in Fig. 4. It was calculated from the GFourier program in the FullProf package. The electron density distribution is helpful in the analysis of ion species distribution in the unit cell. The electron density plots are essential in understanding the interactions between atoms. As described below [Eq. (6)], the inverse Fourier transform of the structure factors is used to calculate electron density $\rho$(x, y, z).

$$\rho_{hkl}(x,y,z) = \Sigma F_{hkl} \frac{e^{-2\pi i(hx+ky+lz)}}{V} \quad (6)$$

where $\rho(x, y, z)$ is the electron scattering density; $F_{hkl}$ gives the structure factor for the respective by Miller plane (*hkl*); and *V* is the unit cell volume.

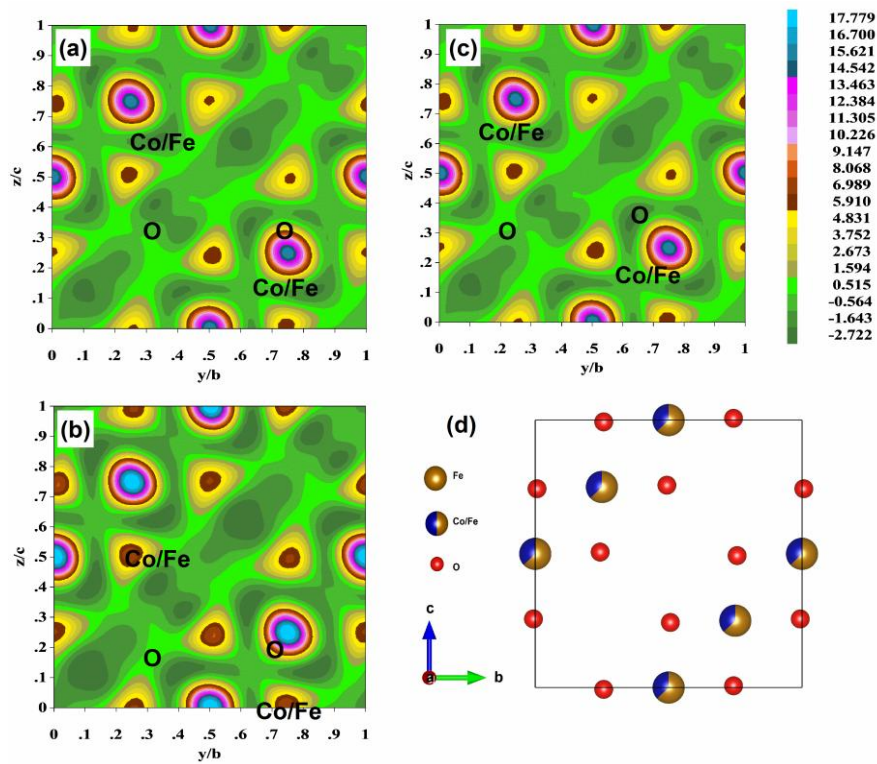

**Fig. 4.** Electron density distribution estimated for CFO nanoparticles annealed at (a) 400 °C, (b) 450 °C, and (c) 500 °C respectively. (d) Shows the cross sectional structure of *yz* plane at *x*=0 intercept.

Here, 2*D*-Fourier contour plots are used to understand the chemical bonding and charge transfer in CFO. We have plotted these from the structure factors obtained from the Rietveld refinement. The contour lines represent the electron density distribution around individual atoms in the lattice structure. Dense contour lines are due the presence of heavy elements. Fig. **4** illustrates the electron density distribution of Co/Fe and O atoms present in the *yz* plane ($x = 0$) of the unit cell. Highly dense circular contour lines around Co/Fe is primarily due to the valence electrons in *d* orbitals. The contour lines surrounding the oxygen are mainly due to the valence electrons present in 2*s* and 2*p* orbitals. The green background present in Fig. 4 is reference level for zero density, while the coloured contour region surrounding Co/Fe and O represents various electron density levels.

### 3.2. FTIR analysis

FTIR spectra of CFO nanoparticles annealed at different temperatures are shown in Fig. 5(b). The CFO possesses a cubic inverse spinel structure with one tetrahedral and one octahedral site with Cobalt and iron at the centre. These ions form metal-oxygen bonds and contribute to the stretching vibrations in tetrahedral and octahedral sites. These vibrations give rise to two major absorption bands, around 460 cm$^{-1}$ and 540 cm$^{-1}$ [44]. The absorption near 540 cm$^{-1}$ is attributed to the stretching vibrations of the metal-oxygen bond (Co/Fe-O) in the tetrahedral site. The more negligible absorption near 460 cm$^{-1}$ is due to the vibration of Fe-O bonds in octahedral sites. The minor absorption in the range of 1000-1500 cm$^{-1}$ is due to the presence of some organic material. The broad absorption peak centred around 3400 cm$^{-1}$ and 1600 cm$^{-1}$ is due to the bending mode of O-H bonds of water [2]. Major IR peaks are summarized in

Table **3**. Thus, the FTIR analysis confirms the presence of the metal-oxygen bonds in the cubic inverse spinel structure of CFO nanoparticles.

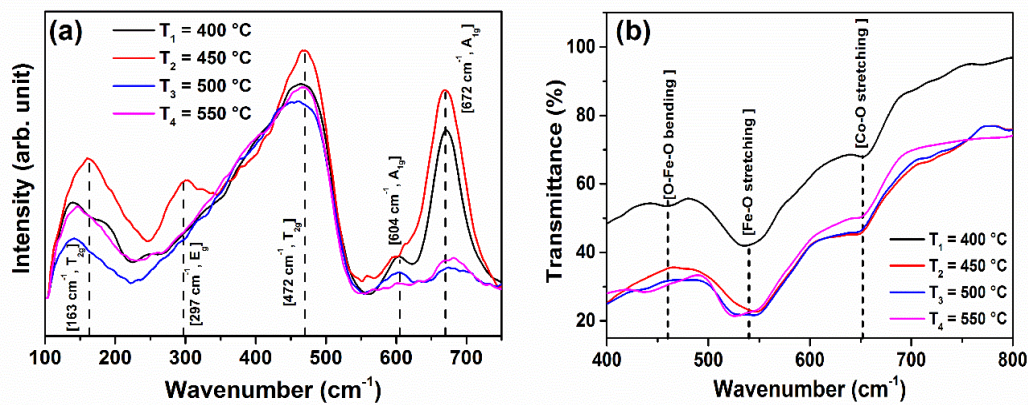

**Fig. 5.** (a) Major Raman modes were observed in CFO nanoparticles annealed at different temperatures. (b) IR transmittance spectra of CFO nanoparticles in the 400-800 cm$^{-1}$ range showing metal-oxygen bonds.

**Table 3:** Major IR peaks are assigned to specific groups of atoms.

| Bond | Type of vibration | Group | Wavenumber (cm$^{-1}$) |
|---|---|---|---|
| Co/Fe-O | Stretching | Octahedral site | ~ 540 |
| Fe-O | Stretching | Tetrahedral site | ~ 460 |
| Co/Fe-O | Bending | Octahedral site | ~ 650 |
| -C=O | Stretching | -COOH | ~ 1740 |
| O=C=O | Stretching | $CO_2$ | ~ 2330 |
| -OH | Stretching | $H_2O$, $C_4H_6O_6$ | ~ 3000 |

### 3.3. Raman analysis

The Raman spectroscopy confirms the presence of inverse spinel structure of CFO nanoparticles. Most of the modes belonging to CFO have been observed in the Raman spectra, as shown in Fig. 5(a). The broadening of these peaks can be attributed to particle size. The bands at 672 cm$^{-1}$ and 604 cm$^{-1}$ correspond to the symmetric stretching of $A_{1g}$ mode (tetrahedral breath mode) of oxygen with respect to Fe$^{+3}$ and Co$^{+2}$ ions. Bands at 472 cm$^{-1}$ and 163 cm$^{-1}$ correspond to the $T_{2g}$ mode (asymmetric stretching and bending), which belongs to the translation motion of the whole tetrahedral structure. The band at 297 cm$^{-1}$ corresponds to $E_g$ symmetric bending of Fe/Co-O bonds [44]. The summary of all results is tabulated below in Table 4.

**Table 4:** Major Raman modes identified in CFO nanoparticles.

| Vibrational modes | Assignment | Raman shift (cm$^{-1}$) | | |
|---|---|---|---|---|
| | | This work | Ref. [45] | Ref. [46] |
| $A_{1g}(1)$ | Symmetric stretching of Fe-O | ~ 672 | 683 | 695 |
| $A_{1g}(2)$ | Symmetric stretching Fe/Co-O | ~ 604 | 617 | 624 |
| $T_{2g}(2)$ | Asymmetric stretching of Fe-O | ~ 560 | 563 | 575 |
| $T_{2g}(3)$ | Asymmetric bending of Co/Fe-O | ~ 472 | 471 | 470 |
| $E_g$ | Symmetric bending of Fe/Co-O | ~ 297 | 300 | 312 |
| $T_{2g}(1)$ | Translation motion of the whole tetrahedral site | ~ 163 | 188 | 210 |

### 3.4. SEM analyses

The SEM images of CFO nanoparticles annealed at 500 °C are shown in Fig. 7 at different resolution levels. It can be seen from the images that the particles were of irregular shape and agglomerated. The particle size was calculated using ImageJ software. The average grain size was around 82 nm, much larger than the size calculated from the XRD analyses due to the polycrystalline in nature. The agglomeration can be a result of the magnetic nature of CFO nanoparticles. The agglomeration and polycrystalline nature can be due to the higher annealing temperature [47]. As discussed earlier, the particle/grain size of nanoparticles depends on the annealing temperature. At higher temperatures, smaller particles fuse to form a larger cluster of particles. Being magnetic nanoparticles, agglomeration seems pretty reasonable for CFO nanoparticles.

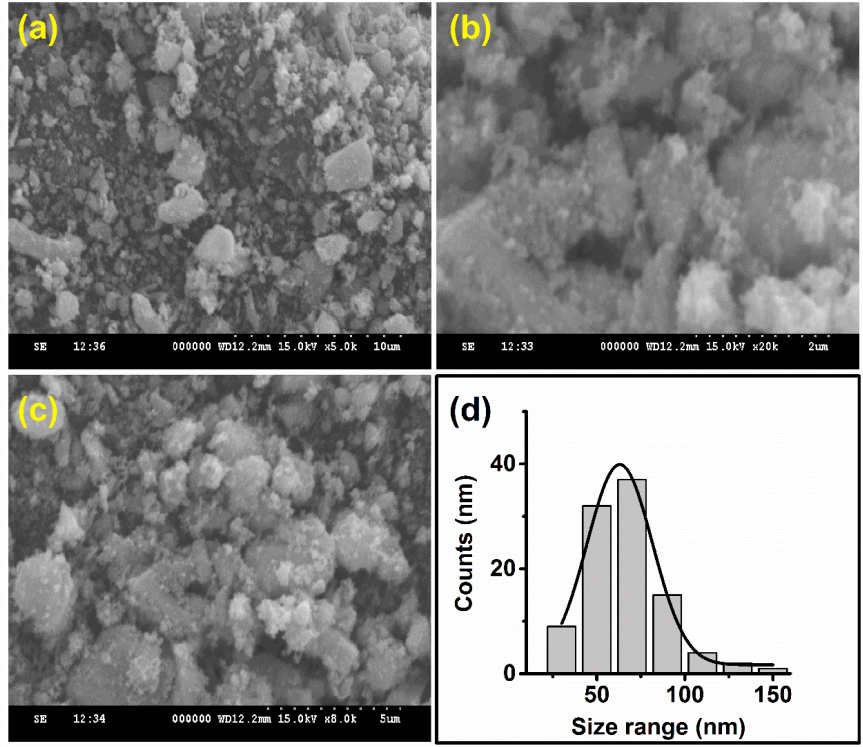

**Fig. 6.** (a-c) SEM micrographs of CFO nanoparticles annealed at 500 °C shown at different resolution level. (d) Distribution of the grain size calculated using ImageJ software.

## 3.5. UV-Vis analysis

The optical property of CFO nanoparticles was studied using UV-Vis diffusion reflectance spectroscopy (DRS). The DRS spectra of CFO nanoparticles annealed at different temperatures are shown in Fig. 7. One can see a sharp reflection edge around 750 nm, illustrating its capability of visible light photocatalyst. The nanoparticle completely absorbs the light in the 400–700 nm region. The ferrites can absorb a large amount of visible light and excite electrons in the O-2$p$ level (valence band) to the Fe-3$d$ level (conduction band) [48]. Kubelka–Munk model was employed to estimate the band gap ($E_g$) of CFO nanoparticles [49]. The absorption coefficient can be estimated by

$$F(R) = \alpha = \frac{(1-R)^2}{2R} \qquad (7)$$

where *F(R)* is the Kubelka-Munk function, R is the reflectance, and $\alpha$ is the absorption coefficient. After that, the band gap ($E_g$) was estimated by using Tauc's equation given by [50],

$$\alpha E = A(E - E_g)^q \qquad (8)$$

It estimates a material's band gap ($E_g$) where $A$ is a constant, and $E$ is the photon energy. The value of $q$ can be two or ½ depending upon the type of transition ($q=1/2$ for direct electronic transitions, $q = 2$ for indirect electronic transitions) [11]. CFO possesses a direct band gap, so the value of $q = ½$. Graph of $(\alpha E)^2$ and $E$ is plotted, and on extrapolating the linear portion of the curve, intersecting on the x-axis gives the bandgap value. Tauc's plot for all samples annealed at different annealing temperatures is shown in Fig. 7 (e-h). The bandgap values for samples $T_1$, $T_2$, $T_3$, and $T_4$ were estimated 1.424, 1.411, 1.398, and 1.384 eV, respectively. These values show a feeble decrement with increasing annealing temperature. The optical property is the key factor in deciding the use of CFO nanoparticles as photocatalysts. The bandgap (~ 1.38-1.42 eV) of CFO nanoparticles lies in the visible region. Thus, it can be concluded from the UV-Vis analysis that CFO nanoparticles can be utilized in the photocatalysis degradation of organic pollutants under visible light.

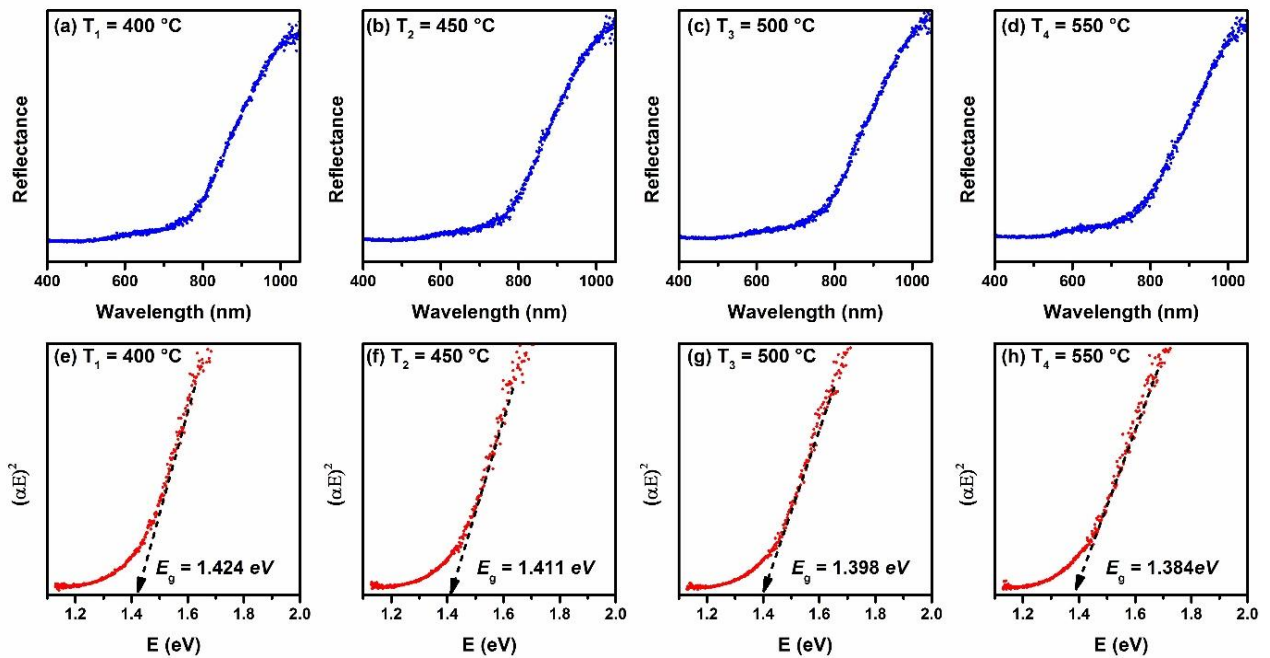

**Fig. 7.** (a-d) UV Vis. DRS spectrographs of CFO nanoparticles annealed at different temperatures. (e-h) estimation of band gap using the Tauc's plot.

### 3.6. Photocatalytic activity

The photocatalytic activity was studied by measuring the degradation rate of MO and MB aqueous solution in the presence of the CFO nanoparticles under visible light. The absorption spectrum of MO shows a significant absorption peak of around ~ 465 nm [51], and that of MB shows around ~ 664 nm. These peaks lie in the visible region and are thus considered for estimating concentration from the UV Vis. spectrum. The photocatalytic activity of CFO nanoparticles (Case 1) was unsatisfactory. It was observed that only 47 % of MO and 12 % of MB dye was degraded even after 150 minutes of irradiation. This may be due to the faster recombination of the electron-hole pair just after generation. So, to restrict this recombination, $H_2O_2$ was employed as an oxidizing agent. It is known for its quick reaction time and degrades into hydroxyl radicals after gaining electrons. These hydroxyl radicals are highly reactive and can improve the reaction rate.

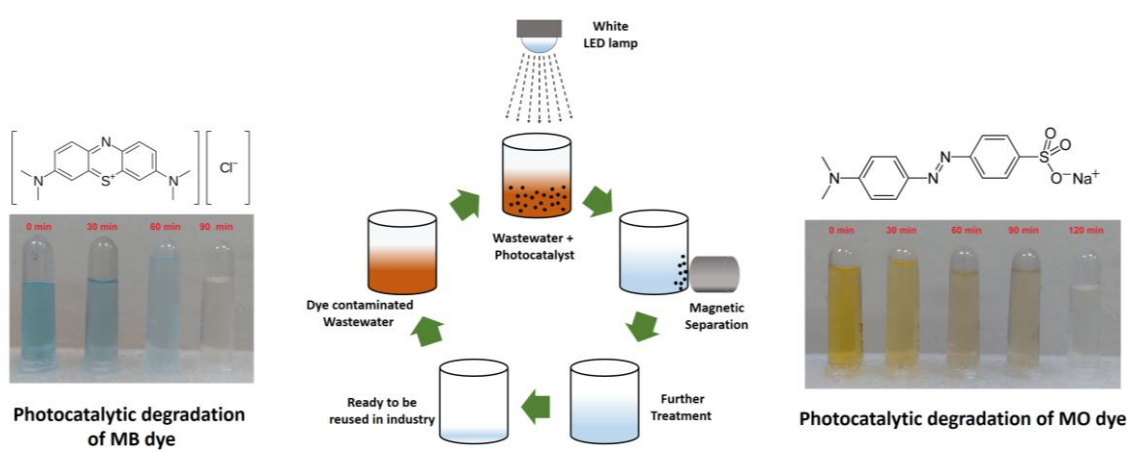

**Fig. 8.** Schematic showing photocatalytic removal cycle of dye-contaminated wastewater.

It is well studied fact that $H_2O_2$ usually requires UV light or metal-ions for activation which cannot be achieved by visible light only [52–54]. The degradation efficiency of photolysis through only $H_2O_2$ under white light was negligible. Thus, to improve the photocatalytic activity of CFO nanoparticles for the degradation of MB and MO dyes, $H_2O_2$ was added (Case 2). It was confirmed from the previous reports that $H_2O_2$ itself shows a poor degradation rate [18,53]. So, the combination of CFO nanoparticles + $H_2O_2$ and LED visible was used to improve the degradation rate of dyes. After 150 minutes of irradiation, most of the dye was found to be degraded. It was observed that within the first 90 minutes of irradiation, 90 % of the dye was found to be degraded. Thus, it confirmed that the photocatalytic activity of CFO nanoparticles was significantly improved with the help of $H_2O_2$ under visible light.

Fig. S2 shows the reduction in the absorbance of the light by MB and MO with respect to time. The concentration value can be calculated using Beer-Lambert's law given by the following equation:

$$\ln\left(\frac{I_0}{I}\right) = A = \epsilon l C \qquad (9)$$

Here, *A, l, ε,* and *C* are absorbance, path length, absorptivity, and concentration, respectively. The degradation efficiency was calculated by measuring the concentration at different time intervals during the experiment. Fig. 9 (a) and (d) show the relative concentration after a specific light irradiation time. It was observed that the relative concentration drops with the increasing time of irradiation, and the degradation efficiency increases with the irradiation time.

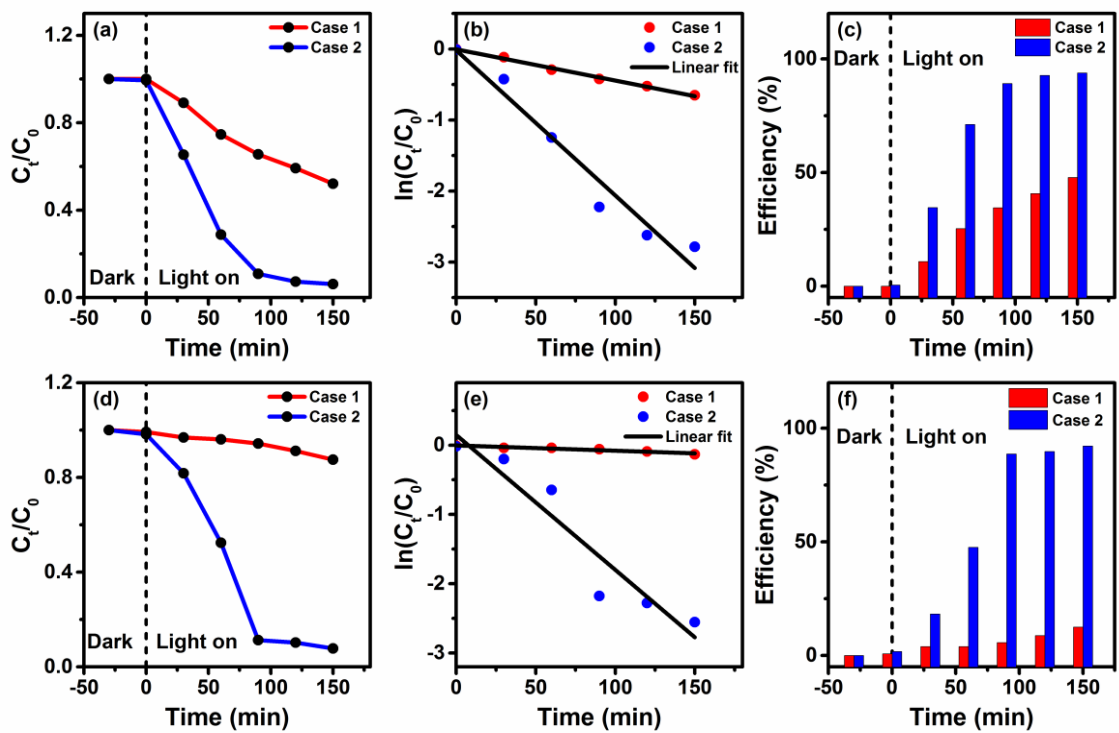

**Fig. 9.** Photocatalytic behavior of CFO nanoparticles: (a) and (d) shows changes in relative concentration over time of MO and MB dye respectively. (b) and (e) shows plots to estimate rate constant based on first order kinematics. (c) and (f) shows the change in the degradation efficiency over time for MO and MB respectively. .

The reaction's first-order rate constant (*k*) was calculated by plotting $ln(C_t/C_0)$ vs. time, fitted linearly to calculate the slope value. The slope value gives the value of the first-order rate constant (*k*). Concentration and degradation efficiency for both cases are given in Table S2 in supporting document. Parameters estimated for this photocatalytic reaction are tabulated in Table 5.

**Table 5** Comparison between rate constant for case 1 and case 2 calculated using first order kinetics.

| Dye | Irradiation time (min.) | The first order rate constant $k$ (min$^{-1}$) | |
|---|---|---|---|
| | | Case 1 | Case 2 |
| Methylene Blue | 150 | $-7.62 \times 10^{-4}$ | $-1.95 \times 10^{-2}$ |
| Methyl Orange | 150 | $-4.40 \times 10^{-3}$ | $-2.04 \times 10^{-2}$ |

### 3.7. Photocatalytic mechanism

The possible mechanism of photocatalytic degradation of MO and MB dyes can be explained through three different pathways. When CFO nanoparticles are irradiated with suitable light ($hv > E_g$), they generate pairs of electrons and holes as described by [55]:

$$CoFe_2O_4 + hv \rightarrow CoFe_2O_4 (e^- + h^+) \tag{10}$$

In the first pathway, the hole produced in the valence band reacts with $H_2O/OH^-$ to generate *OH radicals. These radicals react aggressively with dye molecules to produce degradation products, which are mainly carbon dioxide and water. The following reactions can explain this.

$$h^+ + OH^- \rightarrow {}^*OH \tag{11}$$

$$^*OH + Dye \rightarrow \text{degradation products} (CO_2 + H_2O) \tag{12}$$

In the second possible pathway, electrons produced in the conduction band can with air ($O_2$) and $H_2O_2$ to generate *OH radicals. Again, these radicals react aggressively with dye molecules, converting them into carbon dioxide and water, as the reactions below explain.

$$2e^- + O_2 + 2H^+ \rightarrow H_2O_2 \tag{13}$$

$$H_2O_2 + e^- \rightarrow OH^- + {}^*OH \tag{14}$$

$$^*OH + Dye \rightarrow \text{degradation products} (CO_2 + H_2O) \tag{15}$$

In another possible pathway, $H_2O_2$ may react with $Fe^{+3}$ on the nanoparticle's surface. This will reduce $Fe^{+3}$ to $Fe^{+2}$, as given by Eq. (16). This will again react with $H_2O_2$ to produce *OH.

This radical will degrade the dye into carbon dioxide and water [Eq. (17)]. This process is also known as the Fenton process, and $Fe^{+2}$ produced during the reaction is called the Fenton reagent. This process occurred in the presence of light, so it is also sometimes referred to as the photo-Fenton process.

$Fe^{+3} + H_2O_2 \rightarrow Fe^{+2} + HOO^* + H^+$ (16)

$Fe^{+2} + H_2O_2 \rightarrow Fe^{+3} + OH^- + {}^*OH$ (17)

Thus, the *OH is generated through three different pathways, and it is the main active species responsible for the degradation of the dye molecule.

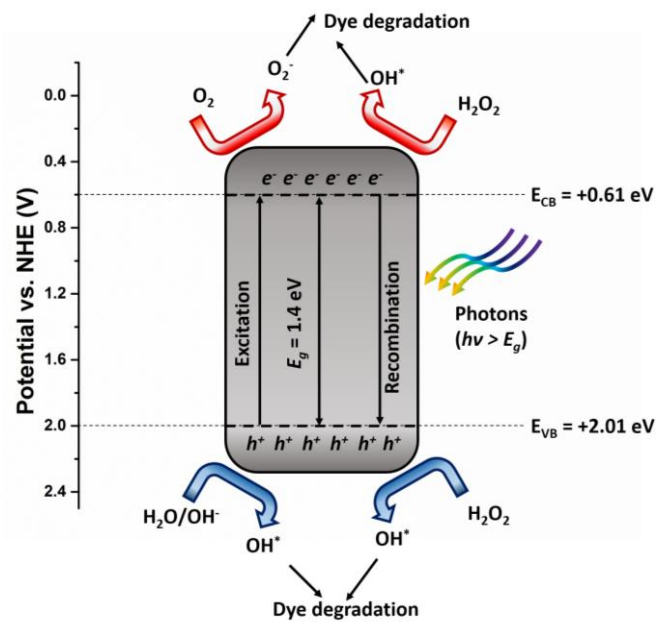

**Fig. 10.** Schematic representation of band energy levels and possible photocatalytic mechanism of CFO nanoparticles.

For better understanding of the mechanism, the band edge energies of valence band ($E_{VB}$) and conduction band ($E_{CB}$) were estimated by following equations,

$$E_{CB} = \chi - E_e - 0.5 E_g$$ (18)

$$E_{VB} = E_g + E_{CB}$$ (19)

Where, $\chi$ and $E_e$ are the absolute electronegativity of the material and free electron energy on the hydrogen scale respectively. The electronegativity of the material was calculated by taking geometric mean of the electro negativities of all the constituent atoms [16,56]. The electronegativity of an atom was estimated by arithmetic mean of its first ionization energy and electron affinity. $E_e$ is the free electron energy relative to hydrogen scale (~ 4.5 eV) [57]. The values of first ionization energy and electron affinity of all atom were taken from the standard database and are tabulated in Table S1.[58,59].

The calculated energy levels for the conduction band ($E_{CB}$) and valence band ($E_{VB}$) are shown in the **Fig. 10**. As photons with sufficient energy is incident on the material, the electron is excited to conduction band and leaves a hole behind in the valence band. This electron-hole pair is responsible for further generation of highly reactive radicals such hydroxyl ($OH^*$) or superoxide ions ($O_2^*$). In case 2, the added $H_2O_2$ inhibits the recombination of this generated charge pair and quickly absorbs it to produce hydroxyl radicals.

### 3.8. Scavenging test

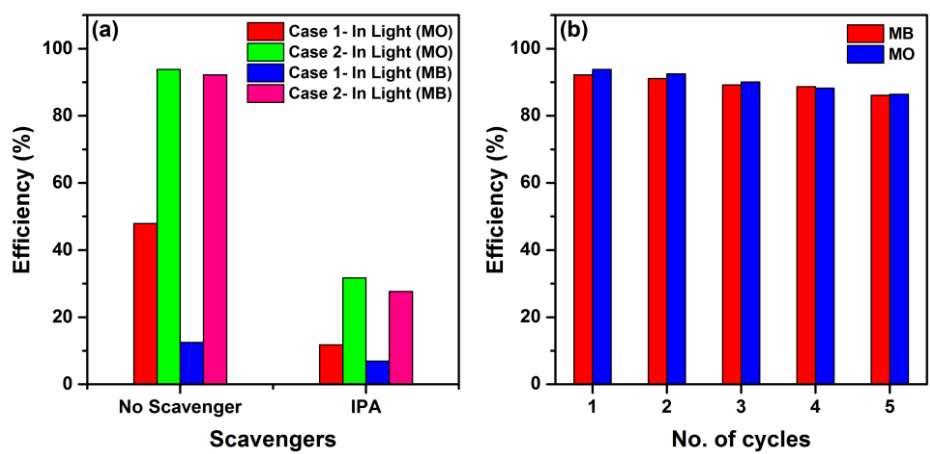

**Fig. 11.** (a) Effect of scavenger on the degradation efficiency of reaction. (b) Recyclabity test for five consecutive cycles performed on CFO nanoparticles for case 2.

To verify the mechanism of the process, the radical trapping test using a scavenger like Isopropyl alcohol (IPA). The process mentioned in section 2.3 was repeated in the absence and presence of IPA. IPA is a a well-known scavenger of hydroxyl radicals (*OH). The degradation efficiency drastically dropped to mere 32% for case 2 for MO and 27% for MB as shown in **Fig. 11**(a). This indicated that the degradation rate was significantly reduced due to the scavenging of *OH due to the addition of IPA. Thus, the *OH radicals are proven to be main active species in this reaction.

### 3.9. Recyclability test

The recycling and reusing capability of any photocatalyst is the main hurdle in the commercialization of this process. Being magnetic, it is very easy to separate and recycle CFO nanoparticles after the reaction using a low-cost magnet. For that, the recyclability and reusability were performed on CFO nanoparticles for case 2. The CFO nanoparticles recovered magnetically and re-processed further for the next cycle. This test was performed for five consecutive cycles for both MO and MB dyes. No significant change in the efficiency was observed after five cycles. A small drop of only 6 % in degradation efficiency was observed in the case of MB and 7% for MO. Efficiency data are shown in detail in Table S4. Therefore, this fluctuation in degradation efficiency can be considered insignificant upto five consecutive cycles. The catalyst was still magnetically recoverable after five cycles. Thus, this test confirms the high structural and chemical stability of CFO nanoparticles. This result is very promising in the commercialization of magnetic CFO nanoparticle-based waste water treatment using photocatalysis.

Therefore, the photocatalytic activity of CFO nanoparticles can be improved by adding an appropriate amount of $H_2O_2$. It increases the production of *OH radicals, the main active species in reaction, and prevents faster recombination of electron-hole pairs generated during irradiation. Thus, the combination of $CoFe_2O_4/H_2O_2$ has shown good photocatalytic activity under visible light. A comparison of this work to some other work is shown in Table 6.

**Table 6**
A comparison of photocatalytic activity results of CFO nanoparticles under visible light.

| Photocatalyst | Dye | Dye concentration (mg/L) | Type of light source | Irradiation time (min.) | Degradation (%) | Ref. |
|---|---|---|---|---|---|---|
| $CoFe_2O_4/H_2O_2$ | RhB | 10 | Visible (30 W LED) | 270 | 90 | [39] |
| $CoFe_2O_4$ | RhB | 6 | Visible (150 W Halide lamp) | 330 | 73 | [60] |
| $CoFe_2O_4/H_2O_2$ | MB | 10 | Visible (Solar simulator) | 140 | 80 | [61] |
| $CoFe_2O_4$ | CV, CR | 25 | Visible (300 W Xe-lamp) | 120 | 82,76 | [62] |
| $CoFe_2O_4$ | MB | 10 | Visible (14 W LED) | 120 | 60 | [38] |
| $CoFe_2O_4/H_2O_2$ | MB, MO | 10 | Visible (12 W LED) | 90 | 90 | This work |

## 4. Conclusion

In conclusion, the CFO nanoparticles were successfully synthesized using the tartaric acid-assisted sol-gel method. The synthesized nanoparticles were characterized by different characterization techniques to analyze the structural properties, optical properties, and photocatalytic activity. The XRD and subsequent Rietveld analysis confirmed the phase purity of synthesized nanoparticles. FTIR and Raman spectrographs further confirm the spinel structure of CFO nanoparticles. The average grain size was around 82 nm, and particles were agglomerated, as observed by the SEM analysis. The UV-Vis DRS was used to analyse the band gap, which was found in a narrow range of ~ 1.38-1.42 eV. This narrow optical band gap lies in the visible region, making these nanoparticles suitable for photocatalytic activity. The photocatalytic activity of CFO nanoparticles was conducted on the aqueous solution of MB and MO dye under low-power white LED (12 W). The photocatalytic activity of nanoparticles under visible light was almost negligible and significantly improved by adding $H_2O_2$ to prevent the recombination of the electron-hole pair. Over 90 % of the dye degradation is observed within 90 minutes.

## Acknowledgement

NP thanks the SVNIT, Surat, institute research fellowship (FIR-D22PH003). This research is supported by the research grant from the Institute seed Money grant 2021–22/DOP/04. The authors are thankful to Ms. Shantilata Sahoo and Dr. D. V. Shah for the UV-visible measurements.